\documentclass[11pt]{article}
\pdfoutput=1 
\usepackage{jheppub}
\usepackage[utf8]{inputenc}

\usepackage{graphicx,ulem}
\usepackage{caption,epigraph}
\usepackage{subcaption}
\usepackage{hyperref}
\usepackage[T1]{fontenc}
\usepackage[export]{adjustbox}
\usepackage[table]{xcolor}
\usepackage{amssymb}
\usepackage{pifont}

\newcommand{\be}{\begin{equation}}
\newcommand{\ee}{\end{equation}}
\newcommand{\bea}{\begin{eqnarray}}
\newcommand{\eea}{\end{eqnarray}}

\DeclareUnicodeCharacter{2212}{-}
\usepackage{multirow, graphicx,amssymb,url,mathrsfs,amsmath}
\usepackage{amsxtra,amstext,latexsym,dsfont,amsfonts}
\usepackage{color,eucal}
\usepackage{float}
\usepackage{colortbl}
\usepackage{multirow}
\usepackage{tikz}
\usetikzlibrary{tikzmark}
\usetikzlibrary{arrows}
\usepackage{tabularx, array}
\newcolumntype{L}[1]{>{\raggedright\arraybackslash}p{#1}}
\newcolumntype{C}[1]{>{\centering\arraybackslash}p{#1}}
\newcolumntype{R}[1]{>{\raggedleft\arraybackslash}p{#1}}
\makeatother
\makeatletter
\newcommand{\myBig}{\bBigg@{1.75}}
\makeatother

\title{\Large Possible enhancement of the superconducting $T_c$ due to sharp Kohn-like soft phonon anomalies}

\author[a]{Cunyuan Jiang,}
\author[b]{Enrico Beneduce,}
\author[a,c,d]{Matteo Baggioli,}
\author[e]{Chandan Setty,}
\author[b]{and Alessio Zaccone}
\affiliation[a]{School of Physics and Astronomy, Shanghai Jiao Tong University, Shanghai 200240, China}
\affiliation[b]{Department of Physics ``A. Pontremoli'', University of Milan, via Celoria 16,
20133 Milan, Italy}
\affiliation[c]{Wilczek Quantum Center, School of Physics and Astronomy, Shanghai Jiao Tong University, Shanghai 200240, China}
\affiliation[d]{Shanghai Research Center for Quantum Sciences, Shanghai 201315, China}
\affiliation[e]{Department of Physics and Astronomy, Rice Center for Quantum Materials, Rice University, Houston, Texas 77005, USA}

\emailAdd{b.matteo@sjtu.edu.cn}
\emailAdd{alessio.zaccone@unimi.it}

\abstract{Phonon softening is a ubiquitous phenomenon in condensed matter systems which is often associated with charge density wave (CDW) instabilities and anharmonicity. The interplay between phonon softening, CDW and superconductivity is a topic of intense debate. In this work, the effects of anomalous soft phonon instabilities on superconductivity are studied based on a recently developed theoretical framework that accounts for phonon damping and softening within the Migdal-Eliashberg theory. Model calculations show that the phonon softening in the form of a sharp dip in the phonon dispersion relation, either acoustic or optical (including the case of Kohn-type anomalies typically associated with CDW), can cause a manifold increase of the electron-phonon coupling constant $\lambda$. This, under certain conditions, which are consistent with the concept of optimal frequency introduced by Bergmann and Rainer, can produce a large increase of the superconducting transition temperature $T_c$. In summary, our results suggest the possibility of reaching high-temperature superconductivity by exploiting soft phonon anomalies restricted in momentum space. 
}

\begin{document} 

\maketitle
\section{Introduction} 
In the early days of classic theories of phonon-mediated superconductivity, such as BCS theory and Migdal-Eliashberg theory, phonons were treated essentially as harmonic oscillators. Things started to change in the late 1970s with the discovery of the ``high-T'' (for that time) superconductivity in Nb-based alloys \cite{Kirzhnits}, at temperatures $T>10$K.

Those materials were characterized by the presence of  structural instabilities connected with quasi-localized lattice (ionic) vibrations, often of the Jahn-Teller type. 
The early models which addressed the influence, and enhancement, of $T_c$ due to these highly anharmonic quasi-local vibrations were developed by Plakida and co-workers, around 1980 \cite{Vujicic_1979,Vujicic_1981}, where the quasi-local vibrations (QLV) are described by means of a pseudospin formalism as two-level anharmonic wells. Within the Eliashberg framework, the contribution to the pairing from these QLVs is taken into account, and shown to lead to a significant enhancement of the $T_{c}$.

This model was subsequently adapted a few years later by Plakida and co-workers \cite{Plakida_1987} to provide an early explanation for the high $T_c$ in the La(Y)BaCuO cuprates discovered in 1986 by Bednorz and M\"{u}ller \cite{Bednorz1986}. 
In this later version of the model, the structural instability in the form of a two-level anharmonic well is attributed to rotational motions of the so-called apical oxygen in the layered perovskite structure, the importance of which has been widely established \cite{Shapiro2021}.
The enhancement of $T_{c}$ is attributed to the fact that highly anharmonic motions, such as those performed by the apical oxygens, are characterized by a large displacement $d$, much larger than the mean squared harmonic displacements $\langle u ^{2}\rangle$. Since the electron-phonon coupling $\lambda$ is approximately proportional to the square of ionic motions, and being $d^{2}/\langle u ^{2}\rangle \gg 1$, this implies a very significant enhancement of $T_c$ brought about by the unstable soft localized vibrations and reflected in the enhanced electron-phonon coupling $\lambda$ with respect to the harmonic one $\lambda_{\text{harm}}$, $\lambda/\lambda_{\text{harm}} \sim d^{2}/\langle u ^{2}\rangle \gg 1$.

While the enhancement of superconductivity caused by localized soft vibrational modes is by now well established for different materials \cite{Shapiro2019}, it may not be enough to explain all the puzzling features of superconductivity in the cuprates, which include $d$-wave symmetry of the paired wavefunction, charge density waves (CDWs) and the effects of magnetic correlations. Furthermore, these early models do not provide a systematic relationship between $T_c$ and phonon damping/softening. Finally these models, while they predict an enhancement of $T_c$ with anharmonic motions, they do not predict the possibility of regimes where instead anharmonicity is detrimental for superconductivity and causes a reduction of $T_c$.

The question of how phonon damping affects superconductivity at a more fundamental level has remained largely unexplored until recently.
In \cite{Setty2020}, the effect of anharmonicity of phonons (both acoustic and optical) has been described at the level of BCS theory. For optical phonons, a non-monotonic dependence of $T_c$ on the single parameter which completely characterized the lattice anharmonicity, i.e. the phonon damping or linewidth, was predicted, with a dome in $T_c$ located at an optimum value of phonon damping. An experimental confirmation of this effect came shortly thereafter in the filled skutterudite ${\mathrm{LaRu}}_{4}{\mathrm{As}}_{12}$, by using electron irradiation to tune the phonon anharmonicity \cite{Shibauchi}.

Damping-assisted enhancement effects have been predicted also as a function of pressure in high-pressure materials \cite{Setty2021}, in qualitative agreement with experiments \cite{Yesudhas}, as a result of ferroelectric soft-mode structural instability leading to the dome in $T_{c}$ at the transition observed in strontium titanate \cite{Setty2022}, and due to glassy effects in cuprates~\cite{Setty2019}.

Phonon softening has been studied in various metallic systems, where the charge correlations soften the phonon spectrum, giving rise to the widely observed ``Kohn anomaly''.
A similar softening, i.e. a dip, in the phonon dispersion relation has been observed and studied, in recent years, also in the cuprates high-T superconductors \cite{LeTacon2014,PhysRev.126.1693,PhysRevResearch.3.013162,PhysRevX.11.041038}. The frequency of these soft modes may go down towards zero even when carrying non-zero momenta (similar to the Peierls instability in 1D materials) at the charge density wave (CDW) wave vector below the ordering temperature $T_{CDW}$, reflecting their origin from CDW order \cite{PhysRevResearch.3.013162,PhysRevLett.30.1144,PhysRevB.13.4258,Wilson}. Early studies examined the role of phonon softening on T$_c$ in low density and strongly coupled superconductors such as strontium titanate~\cite{Appel1969}, as well as transition metals and their Carbides~\cite{Weber1973}.  More recently, the role of phonon softening on the superconducting coupling constant was also studied from first principles in dichalcogenides such as TiSe$_2$~\cite{Mauri2011} and  BiS$_2$~\cite{Yildirim2013}, and simple metals~\cite{Bergara2010}. More specifically, Kohn anomalies were shown to play an important role in determining superconducting properties of many systems, from  hydride superconductors \cite{PhysRevLett.98.117004} to doped graphene \cite{PhysRevLett.105.037002}. Kohn anomalies have been directly associated to the increase of $T_c$ near the lattice instability in NbC$_{1−x}$N$_x$ \cite{blackburn2009superconductivity}.

For the underdoped high-temperature copper-oxide-based superconductors (cuprates), partial softening at the CDW wave vector are observed only below the superconducting critical temperature $T_c$ \cite{PhysRevResearch.3.013162}, in stark contrast to other metallic systems \cite{lorenzo1998neutron,PhysRevB.66.214303}. Using a perturbative approach, this anomalous phonon softening can be explained by a complex interplay among the ubiquitous CDW order, the superconducting (SC) order and thermal fluctuations \cite{PhysRevResearch.3.013162}. Phonon softening has been also reported in several cuprate superconductors outside the superconducting region. Examples of this sort are: La$_{1.875}$Ba$_{0.125}$CuO$_4$ \cite{PhysRevX.8.011008}, La$_{2-x}$Sr$_x$CuO$_4$ (with $x = 0.07, 0.15$) \cite{Reznik2006},
Bi$_2$Sr$_{1.6}$La$_{0.4}$Cu$_2$O$_{6+\delta}$ \cite{PhysRevLett.100.227002},La$_{2-x}$Ba$_x$CuO$_{4+\delta}$ \cite{PhysRevB.78.140511},
HgBa$_2$CuO$_{4+\delta}$ \cite{PhysRevLett.92.197005} (see also \cite{REZNIK201275} for a broader overview).

Let us specify that usually pairing in the cuprates was thought mainly in terms of spin fluctuations \cite{Scalapino1999} and that the normal state from which superconductivity arises is ``strange''. Nevertheless, experimental evidence points to a large electron-phonon coupling in the cuprates \cite{Lanzara2001} with substantial isotope effects~\cite{Lanzara2004,Aiura2008}. Additionally, it was already argued sometime ago by Allen \cite{Allen2001} that traditional arguments against the relevance of phonons for cuprate superconductivity were not watertight. More 
recently, it is becoming increasingly clear that phonons might still explain a variety of cuprate phenomenology~\cite{PhysRevB.104.L140506}, and that fluctuating charge density waves cover a much larger region in the phase diagram, including the strange metallic phase \cite{arpaia2021charge,arpaia2019dynamical} and even the Fermi liquid one \cite{li2022prevailing}. This has led to various speculations about the role of CDW fluctuations for transport properties of strange metals \cite{seibold2021strange,arpaiadiscovering,Baggioli:2022pyb} and about the relation of charge order with quantum criticality and high-temperature superconductivity \cite{PhysRevB.106.155109,arpaia2022signature} as well. Because of these reasons, and aware of all difficulties related to the physics of cuprates, an approach based on Eliashberg theory and phonon-mediated superconductivity might be relevant even for those materials.

While it is natural that the emergence of CDW order changes the phonon behaviour to anomalous softening, then, would it be possible that the anomalous phonon softening can also change the SC properties, and, in particular the $T_{c}$? This is clearly a question of paramount importance for our understanding of high-temperature superconductivity.

Here, using a phenomenological model, we show, by means of a semi-analytical version of Migdal-Eliashberg theory \cite{PhysRevB.101.214502}, that the anomalous phonon softening restricted to sharp region in momentum space, such as the one induced by CDW, can lead to a large enhancement in both the electron-phonon coupling and (under certain conditions) in the superconducting $T_{c}$.

Earlier results \cite{bergmann1973sensitivity} suggest that, in general terms, phonon softening could lead to both an enhancement or a setback of the critical temperature. The main criterion to discern the two situations is linked to the concept of ``optimal frequency", intended as the frequency where the functional derivative $\frac{\delta T_c}{\delta \alpha^2 F(\omega)}$ is maximal. Here, $\alpha^2 F(\omega)$ is the Eliashberg function \cite{PhysRevB.101.214502}. In particular, an enhancement of $T_c$ is expected whenever the softening shifts the weight of $\alpha^2 F(\omega)$ into a more favourable frequency range, i.e., closer to the optimal frequency.

In a different context, recently, a simulation study revealed that the boson-peak-like anomaly in the vibrational density of states (VDOS) of certain metal alloys known as ``strain glasses'' is caused by the phonon softening of the non-transforming matrix surrounding martensitic domains, which occurs in a transverse acoustic branch not associated with the martensitic transformation displacements \cite{ren2021boson}. This means that the phonon softening can support excess modes in the VDOS additional to Debye's law at low frequency, in agreement with the effective theory (``damped phonon'' model) for the VDOS of \cite{PhysRevLett.122.145501,PhysRevResearch.2.013267}. These excess low-frequency vibrational modes, theoretically, should be good to form a lot of strong Cooper pairs, carriers of SC current, and hence causing the $T_c$ of the system to rise. This idea is supported by the experiment on $\mathrm{(Sr_{1-x}Ca_x)_3 Rh_4 Sn_{13}}$ \cite{terasaki2021}, where superconductivity is enhanced by abundant low-energy soft phonons.

In Ref.\cite{PhysRevB.101.214502}, a theory of phonon-mediated superconductivity in strong-coupling amorphous materials based on an effective description of disorder and its damping-like effects on the vibrational spectrum was developed. The theory provides a good analytical description of the Eliashberg function $\alpha^2 F (\omega)$ and therefore can be used in the present work to systematically investigate the effect of phonon softening on superconductivity.

In the following sections, firstly, the theoretical framework leading to the analytical Eliashberg function $\alpha^2 F(\omega)$ is discussed. Then, the damping and softening effects on $\alpha^2 F(\omega)$, on the electron-phonon coupling $\lambda$ and on $T_c$ are shown for acoustic modes. It can be seen that the behaviour of $\lambda$ and $T_c$ as a function of the phonon damping parameter is monotonic: the stronger the damping the smaller $\lambda$ and $T_c$, but the effects are small. Upon introducing they Kohn anomaly-type softening, a peak in the Eliashberg function $\alpha^2F(\omega)$ arises, whose intensity and position correlate positively with the frequency of the soft phonon. It is shown that this peak in $\alpha^2 F (\omega)$ can enhance both $\lambda$ and $T_c$. The softening effect in the optical modes is also studied. Although the analytical expression of $\alpha^2 F (\omega)$ for optical modes is different from the acoustic one, qualitative results are indeed rather similar. A more careful analysis shows that, nevertheless, as expected from the results of \cite{bergmann1973sensitivity}, an increase of $T_c$ due to phonon softening is not always the case. In particular, we will show that there is an ``optimal softening" at which the critical temperature $T_c$ reaches a maximum value. We will provide some heuristic arguments to identify such optimal point.

In the most general setting, this theoretical framework shows the large enhancement of superconductivity arising from soft phonon instabilities, which could be of great importance for the applications of phonon-mediated superconductors on the route towards high-temperature superconductivity. We take an open attitude and we do not exclude that, besides all the aforementioned caveats, our framework might be also relevant for cuprates and strongly-correlated superconductors.

\section{Theory and methods} 
Let us start our analysis by writing the retarded Green's function of a generic excitation with frequency $\omega$, wave-vector $\textbf{q}$ and finite lifetime $\tau$. Assuming rotational invariance, and therefore the Green's function being a function only of $q\equiv |\textbf{q}|$, the latter can be expressed as
\begin{equation}\label{green}
    \mathcal{G}(\omega,q)=\frac{1}{\omega^2-\Omega^2(q)+i \omega\Gamma(q)},
\end{equation}
where $\Omega(q)$ is the renormalized energy and $\Gamma(q)$ the linewidth of the corresponding mode which corresponds to the inverse of its lifetime, $\Gamma=\tau^{-1}$. The form of this Green's function is analogous to the response function of a damped harmonic oscillator in which both the characteristic frequency and the damping term are now wave-vector dependent. The linear in frequency scaling of the imaginary part is dictated by the breaking of time reversal which is equivalent to the presence of irreversible dissipation. In the case of interest, i.e. damped sound waves in an elastic continuum (phonons), the linear in frequency term comes from the fact that the viscous dissipative corrections to the stress are proportional to the strain rate and not the strain itself. The structure in Eq.\eqref{green}, and in particular the linear in frequency scaling of the imaginary part, can be computed explicitly using simple arguments (see for example Section 7.3 in \cite{Lubensky}). For acoustic phonons, in the limit of small wave-vector and finite temperature, we can take the linewidth to be quadratic, $\Gamma(q)=Dq^2$, where $D$ is the damping constant. This diffusive damping is a direct consequence of momentum conservation derivable using effective field theory and hydrodynamics \cite{Lubensky,landau7} and it is independent of whether the solid is amorphous or crystalline \footnote{Here, we avoid complications which arise in the limit of zero temperature where Rayleigh scattering due to harmonic disorder might be dominant in amorphous systems. We also do not consider similar effects on the attenuation constant coming from non-affine motions \cite{doi:10.1063/5.0085199,Baggioli_2022}.}. On the contrary, for optical phonons, the leading term in the relaxation rate $\Gamma(q)$ is wave vector independent, $\Gamma(q)=\Gamma_0$, and is known as Klemens damping \cite{Klemens}. In all our analysis, we take the frequency $\omega$ to be a complex number and the wave-vector $q$ to be a real number. With this choice, a linear excitation contributes to the dynamical response of the system in time as,
\begin{equation}
    e^{-i \omega(q) \,t}\,=\,e^{-i\, \mathrm{Re}[\omega]\,t}\,e^{\mathrm{Im}[\omega] \,t}\,.
\end{equation}
The imaginary part of the frequency determines therefore the lifetime of the excitation, $\tau(q) \equiv - 1/\mathrm{Im}[\omega(q)]$. The requirement of stability then forces $\mathrm{Im}[\omega(q)]<0$ and the Green's function in Eq.\eqref{green} to be analytic in the complex upper plane. Equivalently, this could be derived by imposing causality on the retarded Green's function.

The dispersion relation of the corresponding mode can be obtained by looking at the poles of the retarded Green's function:
\begin{equation}
    \omega^2-\Omega^2(q)+i \omega \Gamma(q)=0.
\end{equation}
In this work, for acoustic modes, we assume
\begin{equation}\label{acu1}
    \Omega(q)=p(q)\left(v\,q-\frac{v}{2q_{\mathrm{VH}}}q^2\right).
\end{equation}
where $v$ is the propagating speed and $q_{\mathrm{VH}}$ is the maximum wave vector corresponding to the van Hove singularity. The function $p(q)$ is a function which controls the softening and which is given by a Gaussian form,
\begin{equation}\label{soft1}
    p(q)=1-\Delta \cdot \mathrm{exp}\left[-\left(\frac{q/q_\mathrm{VH}-\alpha}{\beta}\right)^2\right],
\end{equation}
where $\Delta \in [0,1]$ is a parameter determining the depth of softening, $\alpha \in [0,1]$ defines the wave-vector at which the softening appears, $q_\mathrm{soft}=\alpha q_\mathrm{VH}$, and $\beta \in (0,0.2)$ controls the width of the softening region. At low wave-vectors, Eq.\eqref{acu1} reduces to:
\begin{equation}
    \omega= \tilde{v} q+\dots\,\qquad \tilde{v}=v \left(1-\Delta e^{-\alpha^2/\beta^2}\right),
\end{equation}
which is the typical linear dispersion relation for acoustic modes. For simplicity, we avoid any distinction between transverse and longitudinal modes and consider a single branch with velocity $\tilde{v}$. Moreover, in the following, we will consider always the situation in which:
\begin{equation}
    \Delta e^{-\alpha^2/\beta^2} \ll 1
\end{equation}
such that, at leading order, we can consistently assume $\tilde v=v$. From now on, this distinction will be therefore dropped. Notice also that, by construction, $p(q_\mathrm{VH})=1$.

On the contrary, for optical phonons, we express the renormalized energy as
\begin{equation}
    \Omega(q)=p(q) \,\omega_{\text{max}}
\end{equation}
with $\omega_{\text{max}}$ a $q$-independent frequency as in the Einstein approximation. Once again, to good accuracy, we will always have $p(0)\approx 1$, such that for optical phonons $\Omega(0)=\omega_{\text{max}}$.

The above propagator in Eq.\eqref{green} leads to the typical Lorentzian form for the spectral density
\begin{equation}
    \mathcal{B}(q,\omega)=-\frac{1}{\pi} \mathrm{Im}\mathcal{G}(\omega,q)=\frac{\omega\Gamma(q)}{\pi[(\omega^2-\Omega^2(q))^2+\omega^2\Gamma^2(q)]}.\label{spec}
\end{equation}
Then, the Eliashberg function can be written as (see \cite{Carbotte2003} for a derivation)
\begin{equation}
    \alpha^2F(\boldsymbol{k},\boldsymbol{k'},\omega)\equiv \mathcal{N}(\mu)|g_{\boldsymbol{k},\boldsymbol{k'}}|^2 \mathcal{B}(\boldsymbol{k}-\boldsymbol{k'},\omega), \label{start point}
\end{equation}
where $\mathcal{N}(k_F)$ is the electronic density of states (EDOS) at the Fermi momentum $k_F=\mu$, where $\mu$ is the chemical potential, $g_{\boldsymbol{k},\boldsymbol{k'}}$ is the electron-phonon matrix element and $k,k'$ are the wave-vectors of the paired electrons. According to Engelsberg and Schrieffer  \cite{PhysRev.131.993}, in the case of acoustic phonons as mediators, $g_{q}$ is given by
\begin{equation}
    g_{q}^2=\left|\frac{4\pi Ze^2}{q}\left(\frac{N}{M}\right)^{1/2}\frac{q^2}{q^2+k_s}\right|^2, \label{gq}
\end{equation}
where, $q=|\boldsymbol{k}-\boldsymbol{k}'|$, and $k_s=4e^2 mk_F/\pi$ is the Thomas-Fermi screening length. $Z$ is the atomic number, $N$ the ion density, $M$ the ionic mass, $e$ the electron charge, and $m$ the electronic mass. In the long wavelength limit, we can approximate the matrix element in Eq.\eqref{gq} with the simpler expression $g_q^2=C (vq)^2$ \cite{PhysRev.131.993,PhysRevB.3.3797}, where $C$ is conveniently re-parameterizing all the microscopic information. In the case of optical phonons, $g_q$ can be treated as a constant $g_q^2=C$ \cite{PhysRev.131.993,nature06874,PhysRevLett.115.176401}. In both cases, $C$ determines the strength of the electron-phonon interactions and the size of the matrix element in Eq.\eqref{gq}.

In order to compute the Eliashberg function in Eq.\eqref{start point}, let us first consider a generic function $f(\textbf{q},\omega)$ where $\textbf{q}=\textbf{k}_1-\textbf{k}_2$ and $\textbf{k}_{1,2}$ are wave-vectors defined over the 3D Fermi surface $|\textbf{k}_{1,2} |=k_F$. In 3D, the Fermi surface is the surface of a sphere with radius $k_F$. If the function $f$ is rotationally invariant, then the average over the Fermi surface is equivalent to varying the two angles $\theta,\phi$ while keeping
\begin{equation}
    \textbf{q}^2=2 k_F^2(1+\cos \theta)
\end{equation}
fixed, so the normalized average is just,
\begin{equation}
    \tilde f(\omega)=\frac{1}{S}\int_0^{2\pi} \int_0^\pi\,f(q,\omega)k_F^2 \sin \theta d\theta d\phi
\end{equation}
where,
\begin{equation}
    S=\int_0^{2\pi} \int_0^\pi\,k_F^2 \sin \theta d\theta d\phi
\end{equation}
is the area of the surface of the Fermi sphere $S=4 \pi k_F^2$. Then, we obtain
\begin{equation}
    \tilde f(\omega)=\frac{1}{4\pi}\int_0^{2\pi} \int_0^\pi\,f(q,\omega) \sin \theta d\theta d\phi\,.
\end{equation}
Doing the trivial integration over the angle $\phi$, we then get,
\begin{equation}
    \tilde f(\omega)=\frac{1}{2} \int_0^\pi\,f(q,\omega) \sin \theta d\theta
\end{equation}
which can be re-written as
\begin{equation}
     \tilde f(\omega)=\frac{1}{2}\int_{-1}^{1}f\left(\sqrt{2 k_F^2(1+\xi)},\omega\right)d \xi
\end{equation}
with $\xi=\cos \theta$. Finally, this expression can be further simplified into,
\begin{equation}
     \tilde f(\omega)=\frac{1}{4 k_F^2}\int_0^{4 k_F^2} f\left(\sqrt{\zeta},\omega\right) d\zeta
\end{equation}
where we have re-defined $\zeta=2k_F^2(1+\xi)$.\\
Applying this procedure, the Eliashberg function in Eq.\eqref{start point}, averaged over the Fermi surface (FS) wave-vectors, we obtain 
\begin{equation}
    \alpha^2 F(\omega)=C\int_0^4 k_F^2\, \zeta\, \mathcal{B}\left(k_F \sqrt{\zeta},\omega\right) d\zeta, \label{acoustic result}
\end{equation}
for acoustic modes, and,
\begin{equation}
    \alpha^2 F(\omega)=C\int_0^4 \mathcal{B}\left(k_F \sqrt{\zeta},\omega\right) d\zeta, \label{optical result}
\end{equation}
for optical modes, with $\mathcal{B}(q,\omega)$ the spectral density defined in Eq.\eqref{spec}. Given the Eliashberg function in  Eq.\eqref{acoustic result} and Eq.\eqref{optical result}, we can define the electron-phonon mass enhancement parameter $\lambda$ as \cite{Carbotte2003}
\begin{equation}\label{lala}
\lambda=2 \int_{-\infty}^{\infty}\frac{ \alpha^2 F(\omega)}{\omega}d\omega\,.
\end{equation}
Finally, for the critical temperature, we will follow the approach developed by Allen and Dynes \cite{PhysRevB.12.905} and use the following (Allen-Dynes) formula:
\begin{equation}
      T_c\,=\,\frac{f_1\,f_2\,\omega_{log}}{1.2}\,\exp\left(-\frac{1.04\,(1+\lambda)}{\lambda-u^\star\,-\,0.62\,\lambda\,u^\star}\right)\label{allenformula}
\end{equation}
where 
\begin{equation}
\omega_{log}=\exp \left(\frac{2}{\lambda}\int_{0}^{\infty} d\omega \frac{\alpha^{2}F(\omega)}{\omega} \ln \omega \right)
\end{equation}
represents the characteristic energy scale of phonons for pairing in the strong-coupling limit, $f_1$, $f_2$ are semi-empirical correction factors, as defined in ~\cite{PhysRevB.12.905} and $u^\star$ is the Coulomb parameter.

\section{Results and discussion}
Combining the theoretical framework for the Eliashberg function described above with the Allen and Dynes approach for $T_c$, we are ready to study in detail the effects of phonon softening instability on superconductivity. Since acoustic and optical branches are treated differently, we will first consider the case of acoustic phonons.

\subsection{Effect of damping on superconductivity mediated by acoustic modes}
\begin{figure}[htbp]
    \centering
    \includegraphics[width=\textwidth]{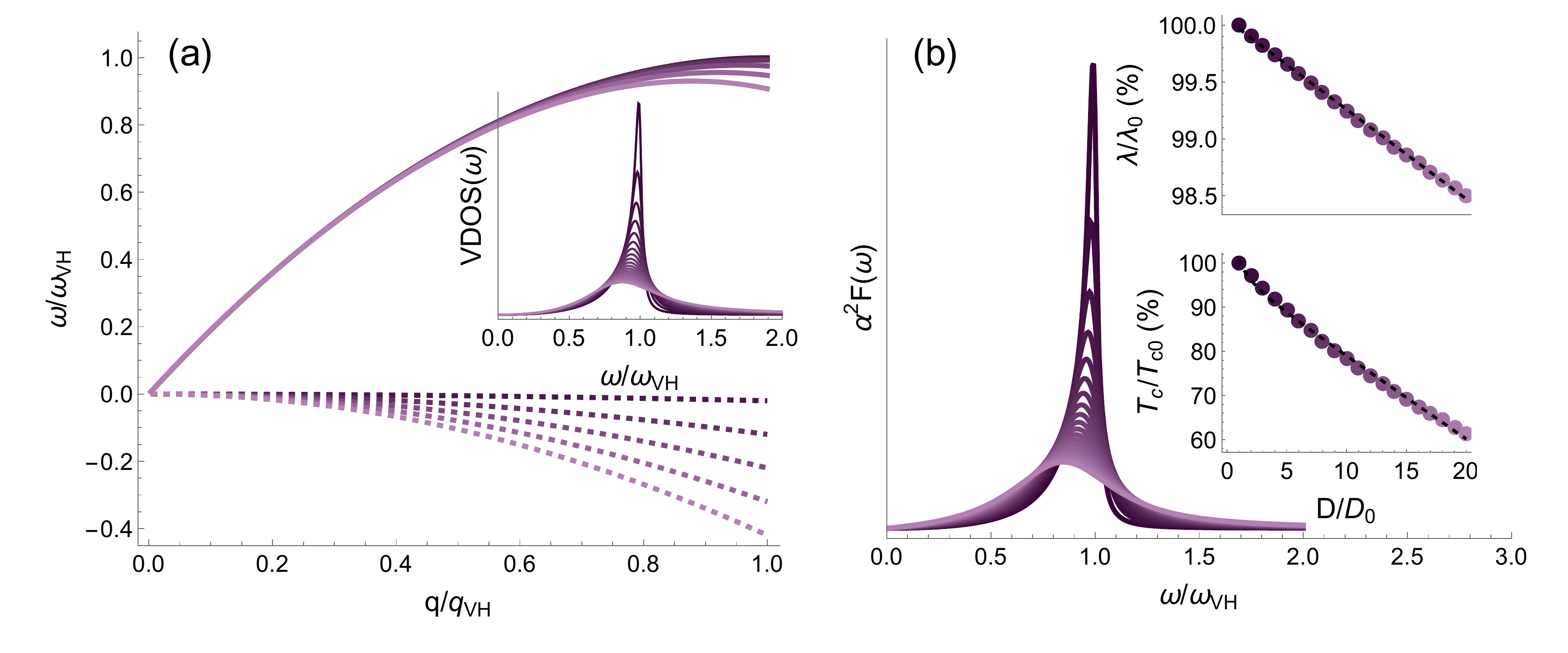}
    \caption{\textbf{(a)}: The real (solid) and imaginary (dashed) part of dispersion relation with different damping constant $D \in [100,2000]$ (from darker to lighter purple). \textbf{(b)}: The Eliashberg function $\alpha^2 F(\omega)$ (in arbitrary units) corresponding to the dispersion relation shown in panel (a). The insets show the coupling constant $\lambda$ and $T_c$ as a function of the damping parameter $D$. $\lambda_0$ and $T_{c,0}$ are the values at the smallest damping $D=D_0=100$. In this figure, the other parameters are fixed to $v=5000$, $q_{\mathrm{VH}}=1$, $k_F=1/2$, $\Delta=0$, $C=0.03$ and $\mu^*=0.1$.}
    \label{fig:plots acoustic D}
\end{figure}

We start our analysis by briefly reviewing the situation in which no softening appears. This will allow us to first describe the effects of phonon damping with no softening in the dispersion relation. The results are shown in Fig.\ref{fig:plots acoustic D}. For simplicity, we will consider only the situation in which the damping $\Gamma(q)$ is small compared to the renormalized energy $\Omega(q)$. Doing so, the real part of the dispersion is always larger than its imaginary part and the mode is a well-defined quasiparticle (even though with a finite lifetime). In this regime, the effect of damping on the real part of the dispersion relation is modest, particularly when $D$ is not too large, as shown in the panel (a) of Fig.\ref{fig:plots acoustic D}. However, the change on the VDOS is more pronounced. As shown in the inset of panel (a), the peak corresponding to the van Hove singularity can be reduced sensitively and smoothed out by increasing the damping $\propto D$. A similar effect can also be found in the Eliashberg function $\alpha^2 F (\omega)$ given by Eq.\eqref{acoustic result}. This "monotonic" change of $\alpha^2 F(\omega)$ is then reflected in a decrease of $\lambda$ and $T_c$ as shown in the inset of panel (b) in Fig.\ref{fig:plots acoustic D}. The peak also moves to lower frequencies since $\mathrm{Re}(\omega)<\Omega$. It should be noticed that the change of $\lambda$ is quite small, only $\approx 1.5\%$. On the contrary, the critical temperature can be reduced by about $\approx 40\%$. This is just a consequence of the highly nonlinear Allen-Dynes relation between $\lambda$ and $T_c$. In a more general setup, whether damping is beneficial or not is a result of the competition between the various factors entering in the Allen-Dynes formula, Eq.\eqref{allenformula}, and depends crucially on the parameters appearing in the dispersion relation of the bosonic mediators \cite{PhysRevB.101.214502}.

\subsection{Effect of softening in the acoustic modes on superconductivity}
After reviewing the case with no softening, we are ready to modify the dispersion relation of the acoustic phonon mediating superconductivity and study the consequent effects. For simplicity, in the rest of the discussion we will keep the damping constant fixed. The softening mechanism is described phenomenologically by the following parameters:
\begin{itemize}
    \item[I.] the energy depth of the softening region which is controlled by the parameter $\Delta$ in Eq.\eqref{soft1};
    \item[II.] the softening position in wave-vector space, which is determined by the parameter $\alpha$ in Eq.\eqref{soft1};
    \item[III.] the width of the softening region, which is parameterized by the parameter $\beta$ in Eq.\eqref{soft1}.
\end{itemize}
In the following, we will analyze the independent effects of these three factors.\\

\begin{figure}[htbp]
    \centering
    \includegraphics[width=\textwidth]{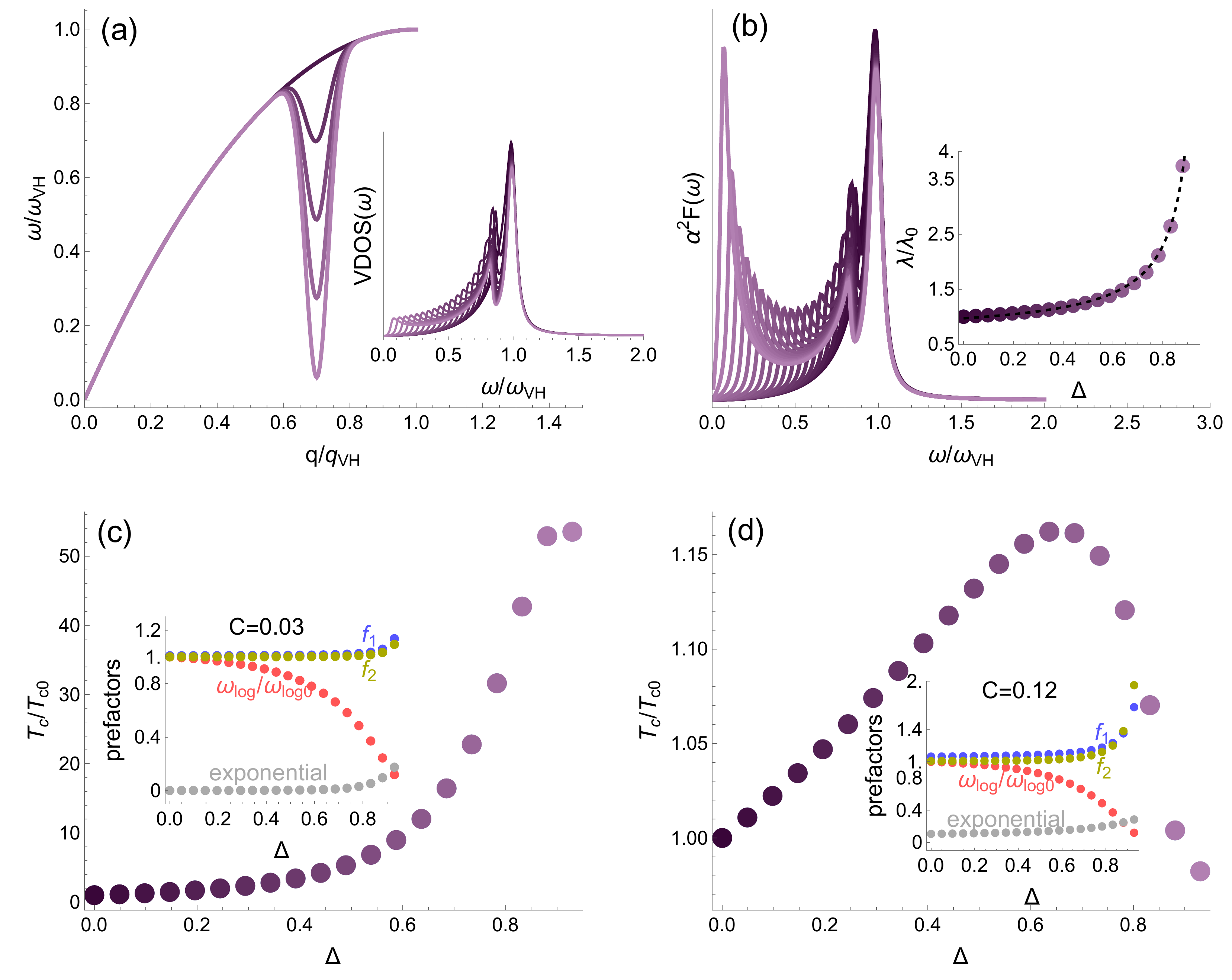}
    \caption{\textbf{(a)}: The real part of dispersion relation for $\Delta \in [0,0.93]$ (from darker to lighter purple). The inset shows the corresponding VDOS. \textbf{(b)}: The Eliashberg function $\alpha^2 F(\omega)$ corresponding to the dispersion relation shown in panel (a). The inset shows the coupling constant $\lambda$ as a function of the softening depth $\Delta$. \textbf{(c)-(d)}: The $T_c$ in function of the softening depth $\Delta$ with $C=0.03, 0.12$ respectively. The insets in panel (c) and (d) show the prefactors $f_1$, $f_2$, $\omega_{\text{log}}$ and the exponential factor in Allen-Dynes theory. $\lambda$ and $T_c$ are normalized by their value without softening, $\lambda_0$ and $T_{c0}$. In the calculations of this figure, $v=5000$, $D=200$, $q_{\mathrm{VH}}=1$, $k_F=1/2$,$\alpha=0.7$ and $\beta=0.05$ are used.}
    \label{fig:plots acoustic Delta}
\end{figure}

We start by considering the effects of the softening region depth $\Delta$. In Fig.\ref{fig:plots acoustic Delta}, the dispersion relation, VDOS, $\alpha^2 F(\omega)$, $\lambda$ and $T_c$ for different values of $\Delta$ are shown. The effect of softening on the dispersion relation is clearly visible in the panel (a) of Fig.\ref{fig:plots acoustic Delta}. In addition to the van Hove peak, both the VDOS and the Eliashberg function $\alpha^2 F(\omega)$ exhibit an additional peak which corresponds to the minimum of the softening region. The peak moves to lower energies by increasing the depth of the softening region $\Delta$, and its intensity increases. Secondary peaks correspond to the inflection points of the dispersion relation due to softening.

 When the depth of the softening region is close to the critical one, i.e. when the energy at the minimum of the softening region approaches zero, the electron-phonon coupling constant $\lambda$, given by $\int_0^\infty d\omega \frac{\alpha^2 F(\omega)}{\omega}$, can be greatly enhanced by the additional softening peak due to the presence of the $1/\omega$ factor therein. This feature is shown in the inset of panel (b) in  Fig.\ref{fig:plots acoustic Delta}. There, $\lambda_0$ indicates the value of the effective coupling in the absence of softening. Whenever $\lambda_0$ is not too large, which corresponds to small values of our parameter $C$ in Eq.\eqref{acoustic result}, the critical temperature $T_c$ can be enhanced greatly and follows a trend which is dominated by the behavior of $\lambda$. This scenario is shown in panel (c) of Fig.\ref{fig:plots acoustic Delta}. The enhancement is reduced only for very large values of the softening depth $\Delta$. In the inset, we show the contribution to $T_c$ coming from the various terms in the Allen-Dynes formula. We observe that all factors appearing in Allen-Dynes formula, Eq.\eqref{allenformula}, apart from the average logarithmic frequency $\omega_{\text{log}}$, grows with the softening depth $\Delta$. We will discuss this in details later. In the opposite situation, i.e. for strongly coupled superconductors, where $\lambda_0$ (and $C$) values are larger, the behavior of the critical temperature is non-monotonic. This case is shown in panel (d) of Fig.\ref{fig:plots acoustic Delta}. First, the normalized critical temperature grows as a function of the softening depth, but then it reaches a maximum at an optimal position and drops down quickly. A first explanation for this behavior can be obtained by looking at the different contributions in the Allen-Dynes formula for $T_c$, Eq.\eqref{allenformula}. In particular, the non-monotonic behavior is simply a result of the competition between the $\omega_{log}$ term which decreases with $\Delta$ and the other terms ($f_1$, $f_2$, and the exponential factor) which, instead, increase with $\Delta$. This implies that, as already noticed in \cite{bergmann1973sensitivity}, softening is not always beneficial for superconductivity.
 
Continuing on these lines, in \cite{bergmann1973sensitivity}, it was argued that there exists an "optimal frequency", $\omega_{\text{opt}}$ given by the maximum in the functional derivative $\frac{\delta T_c}{\delta \alpha^2 F (\omega)}$. Within that theory, softening would be beneficial only if it moves the spectral weight in the Eliashberg function toward frequencies around the optimal frequency. On the contrary, since the Eliashberg function is normalized, if softening moves the spectral weight away from $\omega_{\text{opt}}$, then that would have only a little effect on increasing $T_c$ and the removed weight around $\omega_{\text{opt}}$ would actually lead to a decreasing of the critical temperature.

\begin{figure}[htbp]
    \centering
    \includegraphics[width=\textwidth]{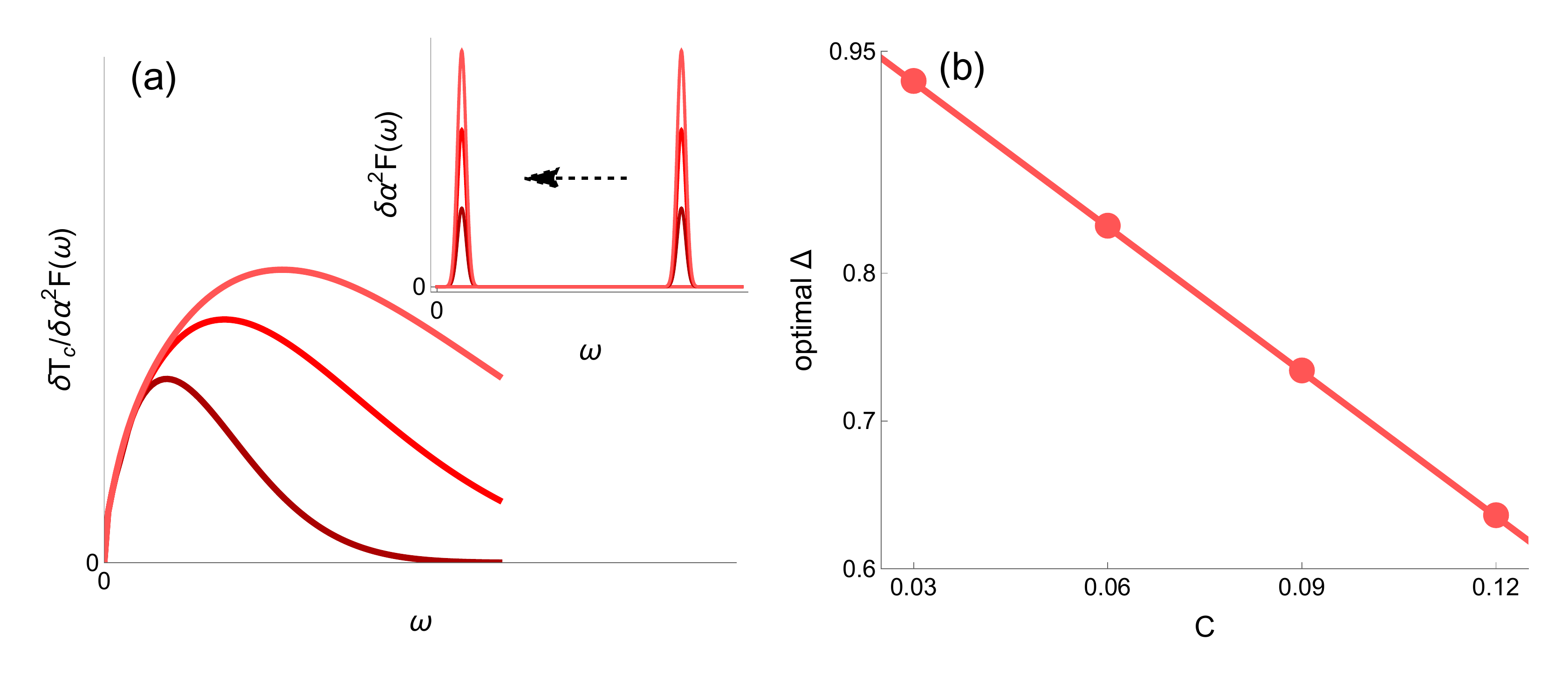}
    \caption{\textbf{(a)}: The functional derivative $\frac{\delta T_c}{\delta \alpha^2 F (\omega)}$ obtained from Allen-Dynes theory by moving an isolated peak of Eliashberg function as shown in the inset. Different colors correspond to different intensities for the artificial peak. \textbf{(b)}: The linear relation of optimal softening $\Delta$ and the constant $C$ obtained from our computations.}
    \label{fig:plots functional derivative}
\end{figure}

\begin{figure}[htbp]
    \centering
    \includegraphics[width=\textwidth]{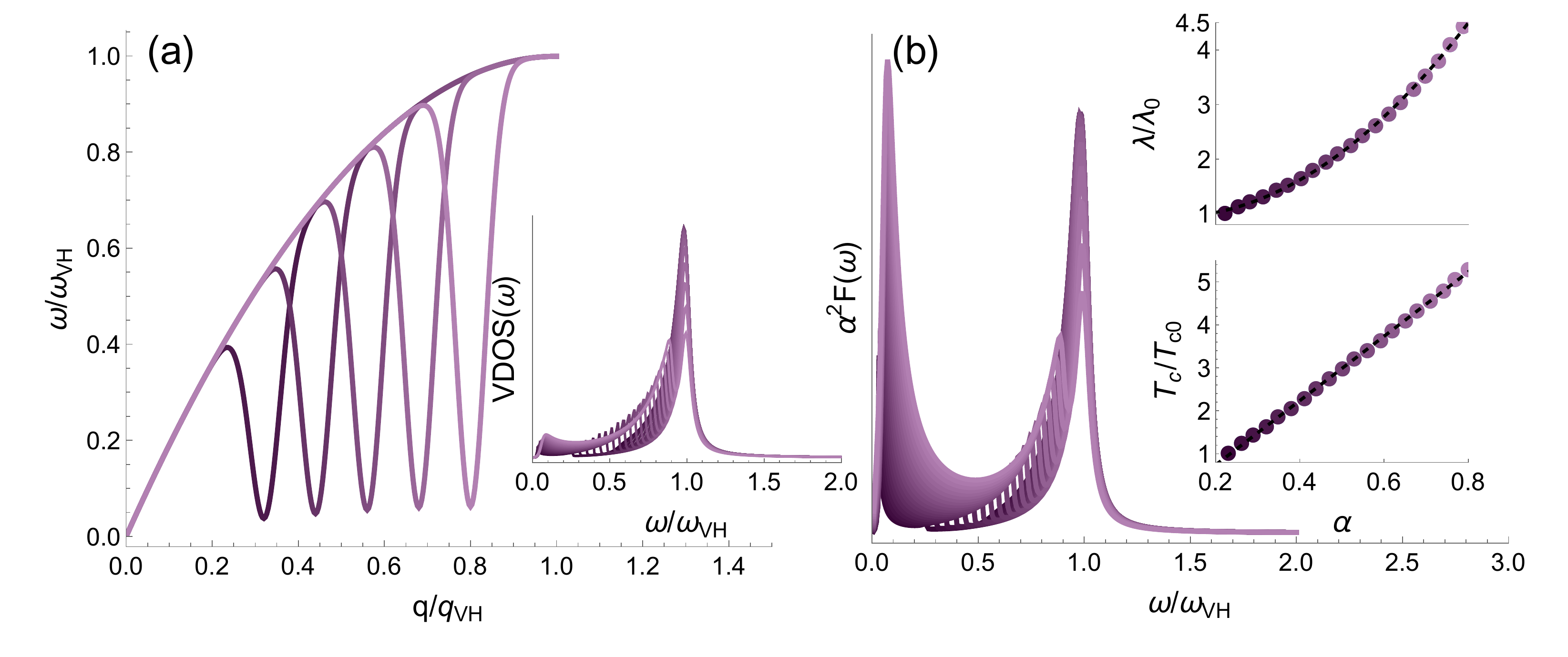}
    \caption{\textbf{(a)}: The real part of dispersion relation for different position of the softening region, $\alpha \in [0.2,0.8]$ (from darker to lighter purple). The inset shows the corresponding VDOS. \textbf{(b)}: The Eliashberg function $\alpha^2 F(\omega)$ corresponding to the dispersion relation shown in panel (a). The insets show the coupling constant $\lambda$ and $T_c$ as a function of $\alpha$. $\lambda$ and $T_c$ are normalized by their value for $\alpha=0.2$. In the calculation of this figure, we have fixed $v=5000$, $q_{\mathrm{VH}}=1$, $k_F=1/2$, $\beta=0.05$, $D=200$, $\mu^*=0.1$ and $C=0.03$.}
    \label{fig:plots acoustic alpha}
\end{figure}

In order to understand the concept of optimal frequency better, we have analyzed a simple situation in which an artificial Gaussian peak is added to the Eliashberg function $\alpha^2 F(\omega)$ (see inset in panel (a) of Fig.\ref{fig:plots functional derivative}) and its energy shifted by hand along the frequency axis. Notice that this situation could be potentially achieved by simply increasing the degree of softening (e.g., making $\Delta$ larger in Fig.\ref{fig:plots acoustic Delta}). We have considered three different cases corresponding to peaks of different intensity. This also could be achieved by simply increasing the parameter $C$ which controls the electron-phonon matrix elements. The functional derivative $\frac{\delta T_c}{\delta \alpha^2 F (\omega)}$ is then calculated numerically as shown in panel (a) of Fig.\ref{fig:plots functional derivative}. From there, we clearly notice the presence of a maximum in the functional derivative as described by \cite{bergmann1973sensitivity}. This implies that the isolated peak gives a strong contribution to the critical temperature around the optimal frequency, but an otherwise negligible contribution if it is away from it. In addition, we notice that the location of the optimal frequency shifts to higher frequencies upon increasing the intensity of the peak in the Eliashberg function, and its shape becomes also more smoothed out. This is consistent with our findings. A larger values of the coupling $C$ corresponds to a peak with stronger intensity which will move the optimal frequency to larger values as in panel (a) of Fig.\ref{fig:plots functional derivative}. A larger optimal frequency implies that a smaller degree of softening, i.e., a smaller $\Delta$, is needed to have a large effect on $T_c$. This is exactly what we observe in panel (b) of Fig.\ref{fig:plots functional derivative}, where the optimal value of $\Delta$\footnote{The concept of optimal $\Delta$ has not to be confused with the concept of optimal frequency, even if obviously connected.} decreases with $C$. This means that a larger depth of softening is required to observe the $T_{c}$ enhancement in weakly-coupled superconductors. Whenever $C$ is very large, and consequently $\lambda_0$ is very large (i.e. strongly coupled superconductors) the optimal $\Delta$ becomes smaller meaning that a smaller degree of softening is needed to increase the critical temperature. This seems to suggest that softening in strongly coupled superconductors can have a stronger effect on $T_c$. However, this is not the case. Comparing panels (c) and (d) in Fig.\ref{fig:plots acoustic Delta}, one immediately realizes that the percentile increase in $T_c$ is extremely small for large values of $C$. As such, a great enhancement of $T_c$ due to softening seems to be possible only when the electron-phonon coupling is small. A strong effect of softening in strongly coupled superconductors is therefore unlikely. In summary, we find that:
\begin{itemize}
    \item A larger electron-phonon coupling ($C$) reduces the optimal softening depth and makes softening easier to meet the optimal frequency conditions. Nevertheless, as shown in panel (d) of Fig.\ref{fig:plots acoustic Delta}, the enhancement will be limited. 
    \item A smaller electron-phonon coupling ($C$) renders the optimal softening depth huge, and in particular close to the value in which the energy of the soft mode approaches zero. Nevertheless, the enhancement due to softening is very large, even before reaching the optimal $\Delta$, as shown in panel (c) of Fig.\ref{fig:plots acoustic Delta}. As such, in this regime, a weak softening in the phonon dispersion can easily increase the critical temperature by a factor $5-10$. On the contrary, this seems impossible (at least withing our simplified toy model) in the case of strong coupling, even when the optimal $\Delta$ is reached. 
\end{itemize}
We continue our study by analyzing the effects of the softening position and width controlled by the parameters $\alpha$ and $\beta$ in Eq.\eqref{soft1}. Fig.\ref{fig:plots acoustic alpha} shows the dispersion relation, VDOS, $\alpha^2 F(\omega)$, $\lambda$ and $T_c$ with different values of $\alpha$ and therefore different positions for the softening region. Here, we keep the depth and width of the softening region constant as shown in panel (a) of Fig.\ref{fig:plots acoustic alpha}. The main result is shown in the inset of panel (b). Moving the softening region to higher values of the wave-vector $q$, closer to the flattening at the Brillouin zone boundary (which leads to a van Hove singularity in the VDOS), both $\lambda$ and $T_c$ grow significantly. This can be explained by the wave-vector dependent prefactor $g^2$ in Eq.\eqref{acoustic result}. The larger the wave-vector at which the softening appears, the greater $g^2$ and hence the larger $\alpha^2 F(\omega)$.

\begin{figure}[htbp]
    \centering
    \includegraphics[width=\textwidth]{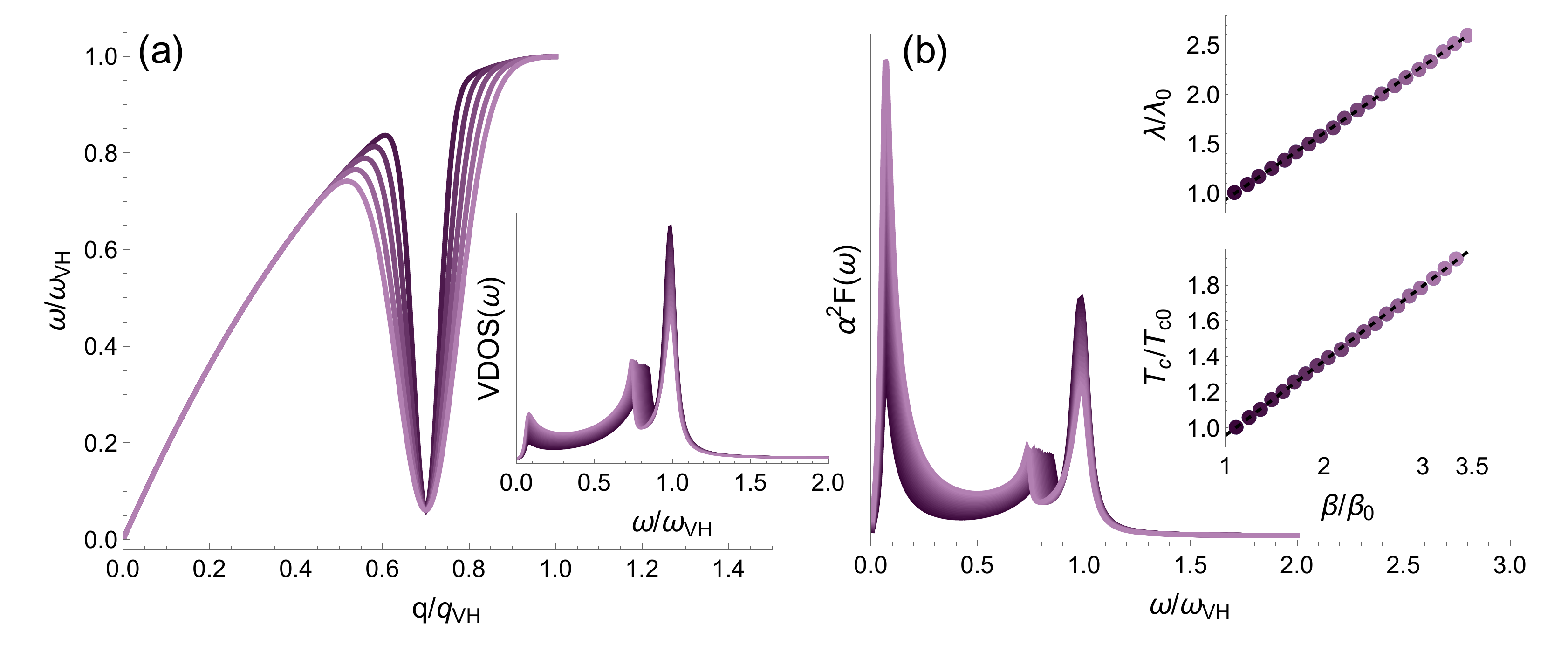}
    \caption{\textbf{(a)}: The real part of the acoustic dispersion relation upon increasing the width of the softening region, $\beta \in [0.03,0.1]$ (from darker to lighter purple). The inset shows the corresponding VDOS. \textbf{(b)}: The Eliashberg function $\alpha^2 F(\omega)$ corresponding to the dispersion relation shown in panel (a). The insets show the coupling constant $\lambda$ and $T_c$ as a function of width parameter $\beta$. $\lambda$ and $T_c$ are normalized by their value at the smallest $\beta=0.03$. In the calculation of this figure, we have fixed $v=5000$, $q_{\mathrm{VH}}=1$, $k_F=1/2$, $\alpha=0.7$, $\Delta=0.93$, $D=200$, $C=0.03$ and $\mu^*=0.1$.}
    \label{fig:plots acoustic beta}
\end{figure}

The width of softening region is also important. Its effects are analyzed in Fig.\ref{fig:plots acoustic beta}. The intensity of the peak corresponding to the minimum of the softening region increases with the width of the latter. By increasing the width of the softening region, both $\lambda$ and $T_c$ increase (see panel (b) of Fig.\ref{fig:plots acoustic beta}). Interestingly, the increase of $T_c$ as a function of $\beta$ is linear.

\subsection{Effect of softening in the optical modes on superconductivity}
As is well known, the properties of optical phonons are quite different from those of acoustic ones. In particular, there are two important differences which need to be emphasized. The first is that, for optical phonons, the lifetime at small wave-vector can be approximately described as a constant, namely $\Gamma(q)=\Gamma_0$, according to Klemens' theory \cite{Klemens}. The second is that the electron-phonon matrix element $g^2$ is also a constant, namely $g^2=\gamma$, as explained in the previous sections. 

\begin{figure}[h!]
    \centering
    \includegraphics[width=0.85\textwidth]{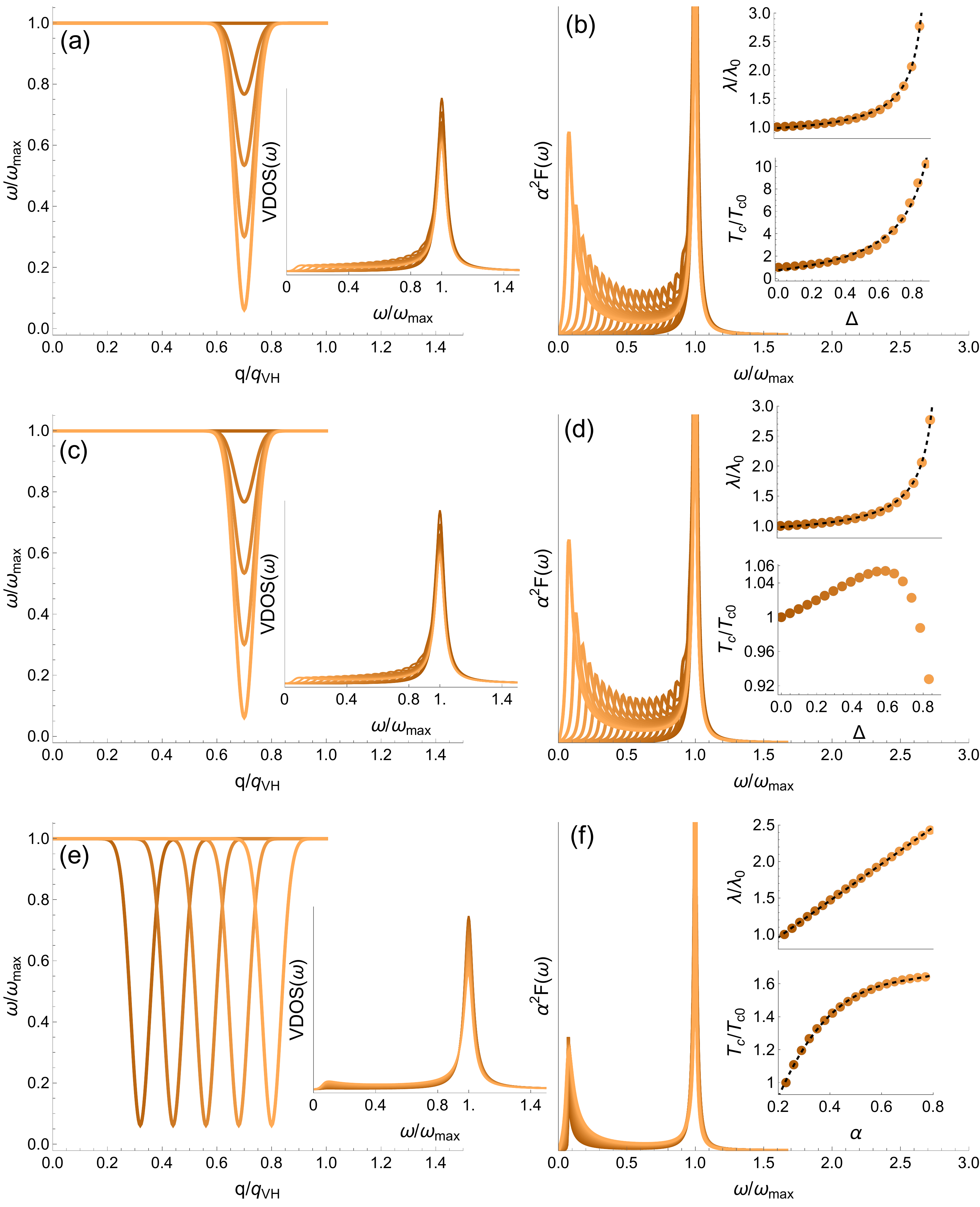}
    \caption{\textbf{Top}: The effects of the softening depth in the case of optical phonons as mediators of the Cooper pairing. The softening parameter $\Delta$ varies in the interval $[0,0.93]$. Here, $\gamma=5 \times 10^5$ and $\alpha=0.7$.
    \textbf{Center}: Same as top panel with $\gamma=10^6$.
    \textbf{Bottom}: The effects of the softening position in the case of optical phonons as mediators. The parameter $\alpha$ varies in the interval $[0.2,0.8]$. Here, $\Delta=0.93$ and $\gamma=5\times  10^5$. In all figures, the other parameters are fixed to $\omega_{max}=3000$, $q_{\mathrm{VH}}=1$, $k_F=1/2$, $\Gamma_0=200$, $\beta=0.05$ and $\mu^*=0.1$.}
    \label{fig:plots optical}
\end{figure}

Let's first consider the effect of the softening depth. The top panel of Fig.\ref{fig:plots optical} shows the dispersion relation, VDOS, $\alpha^2 F(\omega)$, $\lambda$ and $T_c$ for different values of the parameter $\Delta$. The qualitative results are the same as for the case of acoustic phonons with a minor quantitative difference. In particular, at least for the range of parameters explored, the enhancement is smaller than in the case of the acoustic branch (cfr. Fig.2(c)) for the weak-coupling case. This outcome can be interpreted as follows. First, at the minimum of the softening region, the characteristic energy of the optical mode $\omega_{\text{max}}$ is usually significantly larger than for the acoustic case. Therefore, the baseline spectral weight corresponding to the optical mode `\textit{per se}'' is much larger than the one emerging because of softening, rendering the softening effects comparatively weaker. Second, in the case of the acoustic branch, the coupling $g^2$ is wave-vector dependent: $g^2=C(vq)^2$. This means that the modes with larger wave-vector (softened modes) can contribute more to the Eliashberg function $\alpha^2 F (\omega)$ and hence this can enhance the effect of softened modes on the electron-phonon coupling. This is not the case anymore for the optical branch; $g^2$ is wave-vector independent, i.e. a constant, and hence the effects of softened modes are weakened.

The optimal frequency effect discussed in the previous sections for acoustic modes can also be found for optical modes, as shown in the central panels of Fig.\ref{fig:plots optical} using a larger coupling $\gamma$. As shown in the inset of panel (d), the behavior of $T_c$ is again non-monotonic. This implies that the concept of optimal frequency is more general and could be extended also to the case of optical modes. Moreover, it implies that the softening of optical modes is not always beneficial to superconductivity as in the case of acoustic mediators. Finally, let us consider the effect of the position of the softening region. The bottom panels of Fig.\ref{fig:plots optical} show the dispersion relation, VDOS, $\alpha^2 F(\omega)$, $\lambda$ and $T_c$ with different values of the parameter $\alpha$. The enhancement of superconductivity is again not as large as in the case of acoustic mediators. This result can also be interpreted based on the form of the prefactor $g^2$ which is wave-vector independent in this case, and therefore insensitive to the wave-vector related to the location of the softening region. We find that the electron-phonon coupling $\lambda$ grows approximately linearly with $\alpha$, while $T_c$ first grows linearly and then saturates to a constant value. The effects of the softening width have been also analyzed and found to be similar to the acoustic case. Both $\lambda$ and $T_c$ can be enhanced in a linear way as the softening region becomes wider. To avoid clutter, we have decided not to show them.

\section{Outlook}
In this work, we have considered the effects of phonon softening, localized in a small range of wave vectors, on the critical temperature of superconducting materials using the Eliashberg formalism. To simplify the mathematical treatment, we have parameterized the phonon softening as a Gaussian-shaped dip in the dispersion relation of the phonon modes. We have considered both the cases of acoustic and optical modes and analyzed the effects of softening in terms of three parameters: the energy depth of the dip in the phonon dispersion relation, its width and its location in wave-vector space.

We have observed that, under optimal conditions, the softening can strongly enhance both the electron-phonon coupling and the superconducting critical temperature. Our results suggest that for materials in which the electron-phonon coupling is larger, a smaller degree of softening is needed to reach the optimal effect on $T_c$ but the latter is rather limited. On the contrary, for systems with a smaller electron-phonon coupling, a larger degree of softening is required but the final effect is much larger. It would be interesting to improve our analysis and quantify this statement more precisely. Finally, as anticipated by Bergmann and Rainer \cite{bergmann1973sensitivity}, we have explicitly verified that phonon softening does not always lead to an enhancement of the critical temperature as our conclusions agree with the concept of ``optimal frequency'' introduced in \cite{bergmann1973sensitivity}.

In this work, we have restricted our study to a phenomenological model of dispersion relations with softening for the bosonic mediators and to a qualitative analysis that has nevertheless already revealed several interesting features. More in general, the interplay and competition of charge order (or other types of order) with superconductivity is an extremely important question in condensed matter physics. Our calculations demonstrate that these sharp soft phonon anomalies induced by charge order can lead to a strong enhancement of electron-phonon coupling and of the superconducting critical temperature compatible with high-temperature superconductivity despite the softening being restricted in momentum space. 

The enhancement of the superconducting temperature due to the onset of charge order or to giant phonon softening has been observed in a large class of materials \cite{PhysRevLett.109.097002,PhysRevB.73.024512,Houben2020,PhysRevB.102.024507,PhysRevLett.119.087003,PhysRevB.93.184512,Gruner2017,Park2021,doi:10.1073/pnas.2109406119} and theoretically rationalized in the past \cite{PhysRevB.92.125108,PhysRevLett.75.4650,Fanfarillo_2016}. 
The same type of phonon softening described here is present in metallic hydrogen at high pressures \cite{hydrogen}, at temperatures compatible with the onset of superconductivity.

Here, we provided a simple but concrete verification of this possibility in the context of boson-mediated Eliashberg theory. We also notice that our results are not at odd with several other situations (e.g., \cite{barber2021suppression,PhysRevB.98.064513,Yu2021,Chikina2020}) in which the competition between CDWs and superconductivity has been observed to be detrimental for the critical temperature $T_c$. In fact, as verified by our computations, and already envisaged by the seminal work by Bergmann and Rainer \cite{bergmann1973sensitivity}, softening does not always lead to an increase of $T_c$. In a nutshell, whether or not CDW (and phonon softening in general) boosts, competes or coexists with superconductivity is still unknown, as there are experimental measurements which support both scenarios. Our theoretical model justifies this landscape of possibilities by mean of the "optimal softening" criterion. It would be important to provide a more physical explanation for this criterion. From a material design perspective, it is fundamental to realize this concept in order to identify materials in which the effect of softening can be highly beneficial for superconductivity.

For this purpose, it would be helpful to consider scenarios in which the softening of the phonons can be controlled by means of external pressure and strain as performed in \cite{PhysRevB.97.020503,doi:10.1126/science.aat4708}. There, an inverse correlation between the strength of the CDW and the superconducting temperature $T_c$ has been observed. A possible explanation for these cases might be given by the fact that the softening regime appears far from the optimal frequency in the sense of Bergmann and Reiner \cite{bergmann1973sensitivity}. From a model building perspective, it would be interesting to understand and describe how the optimal softening/frequency changes under external mechanical deformations to verify whether the above picture can be confirmed.

An interesting class of materials to which our framework could be applied include NbSe$_2$ and BaNi$_2$As$_2$ in which soft phonons are widely observed above the critical superconducting temperature but their origin with CDW instabilities is not fully understood yet \cite{PhysRevB.86.155125,souliou2022soft,PhysRevB.106.144507,PhysRevLett.107.107403}.

Finally, it would be interesting to expand our analysis to more realistic scenarios, e.g. $d$-wave superconductivity (see for example \cite{PhysRevB.54.16216}), and to attempt a more quantitative analysis of the problem. This would allow us to ascertain whether our results might be relevant to Kohn-type soft phonon anomalies observed in the cuprates, which are closely associated with the ubiquitous charge density wave (CDW) order \cite{arpaia2021charge}. Interestingly, in YBa$_2$Cu$_3$O$_{6.6}$  \cite{LeTacon2014} (see also \cite{doi:10.7566/JPSJ.90.111006}), giant phonon softening was observed below the superconducting critical temperature. This is quite different from our setup, in which phonon softening is considered in the normal phase above the superconducting critical temperature\footnote{We thank Matthieu Le Tacon for explanations about this point.}. Nevertheless, it would be interesting to investigate whether our calculations bear any relations with the findings of \cite{LeTacon2014}. More in general, cuprate superconductors are often associated to phonon softening due to charge order \cite{PhysRevX.8.011008,Reznik2006,PhysRevLett.100.227002,PhysRevB.78.140511,PhysRevLett.92.197005,REZNIK201275}, and despite their ``strange'' features, are good candidates to test our model.\\

Undoubtedly, a more quantitative estimate of the effects of phonon softening on the critical temperature would be very desirable. At this stage, we are only able to come up with an estimate of the percentile increase of $T_c$ as a function of softening. In order to provide a more precise prediction for the enhancement, several difficulties have to be overcome. First, realistic values for the input parameters of the model have to be obtained from experiments or simulations. Second, the validity of our phenomenological softening-function must be validated using the actual phonon dispersions in concrete materials. Third, one should verify that pairing, and the corresponding Eliashberg function, are dominated by the phonons which soften and other modes have only a subdominant effect on $T_c$.
We leave these tasks for future research.

\section*{Acknowledgements}
This work is dedicated to the memory of K. Alex M\"{u}ller, who passed away while this manuscript was prepared, and whose work on soft modes and superconductivity has been a constant inspiration for one of us (A.Z.).
We would like to thank A.~Balatsky, C.~Pepin, M.~Le Tacon and Y.~Ishii for fruitful discussions and comments on a preliminary version of this draft. 
M.B. acknowledges the support of the Shanghai Municipal Science and Technology Major Project (Grant No.2019SHZDZX01) and the sponsorship from the Yangyang Development Fund. M.B. would like to thank IFT Madrid, NORDITA, GIST, APCTP and Chulalongkorn University for the warm hospitality during the completion of this work and acknowledges the support of the NORDITA distinguished visitor program and GIST visitor program. 
A.Z. gratefully acknowledges funding from the European Union through Horizon Europe ERC Grant number: 101043968 ``Multimech'', and from US Army Research Office through contract nr. W911NF-22-2-0256.

  \bibliographystyle{JHEP}
\bibliography{references}

\end{document}